\documentclass[aps,prb,twocolumn]{revtex4}

\usepackage{graphicx}
\usepackage{dcolumn}
\usepackage{amsmath}
\usepackage{color}
\usepackage{soul}
\usepackage{gensymb}

\newcommand{\TB}{TbMnO$_3$ }

\begin{document}
	
	\title{Elastic and magnetoelastic properties of \TB single crystal by nanosecond time resolved acoustics and first-principles calculations}
	
	\author{P. Hemme$^1$}
	\author{C-H. Li$^{2,3}$}
	\author{P. Djemia$^2$}
	\author{P. Rovillain$^4$}
	\author{S. Houver$^1$}
	\author{Y. Gallais$^1$}
	\author{A. Sacuto$^1$}
	\author{H. Sakata$^5$}
	\author{S. Nowak$^6$}
	\author{B. Baptiste$^7$}
	\author{E. Charron$^4$}
	\author{B. Perrin$^4$}
	\author{L. Belliard$^4$}
	\author{M. Cazayous$^{1}$}\thanks{corresponding author : maximilien.cazayous@u-paris.fr}

	\affiliation{$^1$Laboratoire Mat\'eriaux et Ph\'enom$\grave{e}$nes Quantiques, Universit\'e de Paris, UMR 7162 CNRS, 75205 Paris Cedex 13, France\linebreak
		$^2$Laboratoire des Sciences des Procédés et des Matériaux UPR-CNRS 3407, Universit\'e Sorbonne Paris Nord, Alliance Sorbonne Paris Cité, Villetaneuse, 93430, France\linebreak
		$^3$School of Materials Science and Engineering, Beijing Institute of Technology, Beijing, 100081, China\linebreak 
		$^4$ Institut des Nanosciences de Paris, Sorbonne Universit\'e, CNRS UMR 7588, 4 place Jussieu, 75005 Paris, France\linebreak
		$^5$Department of Physics, Tokyo University of Science, 1-3 Kagurazaka Shinjyuku-ku Tokyo, Japan 162-8601\linebreak
		$^6$UFR de Chimie, Universit\'e de Paris, 15 rue Jean Antoine de Ba\"if, 75013 Paris, France\linebreak
		$^7$Institut de Min\'eralogie, de Physique des Mat\'eriaux et de Cosmochimie, CNRS UMR 7590, Sorbonne Universit\'e, 75005 Paris, France}
	
	\begin{abstract}
		Time resolved pump and probe acoustics and first-principles calculations were employed to assess elastic properties of the TbMnO$_3$ perovskite manganite having orthorhombic symmetry. Measuring sound velocities of bulk longitudinal and shear acoustic waves propagating along at least two different directions in the high symmetry planes (100), (010) and (001), provided a powerful mean to selectively determine the six diagonal elastic constants C$_{11}$= 227 GPa, C$_{22}$= 349 GPa, C$_{33}$= 274 GPa, C$_{44}$= 71 GPa, C$_{55}$= 57 GPa, C$_{66}$= 62 GPa. Among the three remaining off-diagonal ones, C$_{23}$= 103 GPa was determined with a 
		bissectrice direction. Density functional theory calculations with colinear spin-polarized provided complementary insights on their optical, elastic and magnetoelastic properties.         
	\end{abstract}
	
	\maketitle
	
	
Multiferroics are well known materials that display simultaneously ferroelectric and magnetic properties.\cite{Eerenstein} Beside their exciting physics, strong magnetoelectric coupling, present in some of them, has opened up a whole field of applications for new spin-based devices\cite{Bea, Spaldin}. 
Several multiferroics own an original proccess to induce ferroelectricity from a magnetic state, the so-called improper multiferroics such as the perovskite manganites TbMnO$_3$.\cite{Khomskii} The ferroelectricity in \TB has its origin from the magnetic exchange striction \cite{Sergienko2006b, Picozzi2007} or from a spin order breaking the spatial inversion symmetry via the spin-orbit interaction.\cite{Kenzelmann2005, Mostovoy2006, Hu2008, Mochizuki2009} 
The novel couplings between microscopic degrees of freedom such as spin and charge is the main reason why TbMnO$_3$ is one of the most intensively investigated  magnetoelectric manganite among the frustrated magnets. The strength of the magnetoelectric coupling gives rise to unusual dynamical effects like electromagnons, spin waves that are excited by the electric-field component of  light\cite{Pimenov2006, Senff2007, Takahashi2008, Aguilar2009, Rovillain2010}.
One of the two electromagnons observed in TbMnO$_3$ has already been explained as a zone-edge magnon activated purely by the magnetostriction mechanism. \cite{Aguilar2009, Stenberg2009, Mochizuki2010a} 
Yet, the multiferroic properties of these compounds exist only at very low temperature with a weak polarization, which limits application perspectives. However, a possible start of application has been demonstrated using electromagnons to modify the  atomic-scale magnetic structure of TbMnO$_3$ with THz optical pulses \cite{Kubacka2014}. In addition, 
strain engineering has emerged as a powerful means for tuning the multiferroic mechanisms of perovskite oxide thin films.\cite{Shimamoto}
The determination of the magnetostriction and magnetoelastic coefficients, and consequently the knowledge of the elastic coefficients are essential to be able to quantify the interactions that are at work in these complex couplings.
To the best of our knowledge, there is no experimental determination of the TbMnO$_3$ elastic coefficients. The few works on the subject are theoretical,\cite{Choithrani2011} one employing a shell model with transferable pairwise interionic interaction potential.\cite{Choithrani2009} In order to remedy this shortcoming, we applied the pump-probe time-resolved acoustics method we recently developed for the rhombohedric multiferroic BiFeO$_3$\cite{Pierre2021} to TbMnO$_3$ which has 9 independent elastic constants. This approach has been extended to the magnetoelastic properties which knowledge is crucial for this type of compound.  

In this work, sound velocities of both longitudinal and transverse acoustic modes traveling along different directions in single crystals TbMnO$_3$ have been measured using 
an acoustical pump-probe experiment at the nanosecond time-scale. To date, no other experimental technique allows to do this on a material of which only small samples exist. To meet the challenge of performing a successful analysis of these data to identify the maximum of elastic constants, we used the Christoffel equation and elastic constants calculated by the density functional theory (DFT). We have been able to determine a set of initial theoretical velocities. Then, among all nine independent C$_{ij}$ elastic constants of TbMnO$_3$, seven have been selectively measured along high symmetry directions of the lattice. In addition, we have tracked unambiguously each acoustic wave by identifying simultaneously their sound velocities and their different polarizations. Lastly, the anisotropic magnetostriction and magnetoleastic coefficients of TbMnO$_3$ has been simulated by DFT and compared to the calculated ones of BiFeO$_3$.

We have investigated three different planes  [$\bf{b},\bf{c}$], [$\bf{a},\bf{c}$] and [$\bf{a},\bf{b}$] of highest symmetry, all oriented from the same TbMnO$_3$ single crystal grown by floating-zone method.  TbMnO$_3$ becomes antiferromagnetic below the N\'eel temperature T$_N$=42~K.\cite{Quezel1977} In this phase the Mn magnetic moments form an incommensurate sinusoidal wave with a modulation vector along the $b$ axis. The ferroelectric order appears below T$_C$=28~K at the magnetic transition from incommensurate to commensurate order where the spin modulation becomes a cycloid.\cite{Kimura2003} 
The crystal axes and the lattice parameters have been determined for each samples using a X-ray diffractometer with a 4-axis goniometer.  
TbMnO$_3$ crystallizes in the orthorhombic symmetry (P\textit{nma}) with lattice parameters equal to $a=5.833\pm 0.080~ \rm \mathring{A}$, $b=7.429 \pm0.050~ \rm \mathring{A}$, $c=5.301\pm0.050~\rm \mathring{A}$ and a mass density $\rho$=7.51$\pm0.2$ g/cm$^{3}$. They compare well to reported values $a=5.838~ \rm \mathring{A}$, $b=7.402~ \rm \mathring{A}$, $c=5.293~\rm \mathring{A}$ with a mass density $\rho$=7.603 g/cm$^{3}$.\cite{Alonso2000}
The crystals have been polished and an aluminium thin film of 75 nm has been evaporated on the surface. For our pump and probe optical method, the Al allows to enhance the acoustic waves generated in TbMnO$_3$ by the pump and  the reflectivity signal from the surface displacements detected by the probe. The thickness of Al is very small compared to the wavelengths of the acoustic waves involved in our experiment, and therefore should not disturb significantly the dispersion curves of acoustic waves and the propagation speeds.

Acoustical pump-probe method has been implemented to measure sound velocity within the plane of three differently oriented single crystals (100), (001) and (010), and thus to determine elastic constants.
The source is a  mode-locked Ti:sapphire 
The pump probe time-delayed is tuned using a mobile reflector with a maximum amplitude of 12 ns. The surface displacements are measured by a Michelson interferometer detecting the phase  of the reflectivity changes.
A more detailed description of our pump‐probe experiment can be found in the supplementary material and in previous works.\cite{Belliard, Amziane, Xu}


\begin{figure}[tb!]
	\begin{center}
		\includegraphics[width=4.5cm]{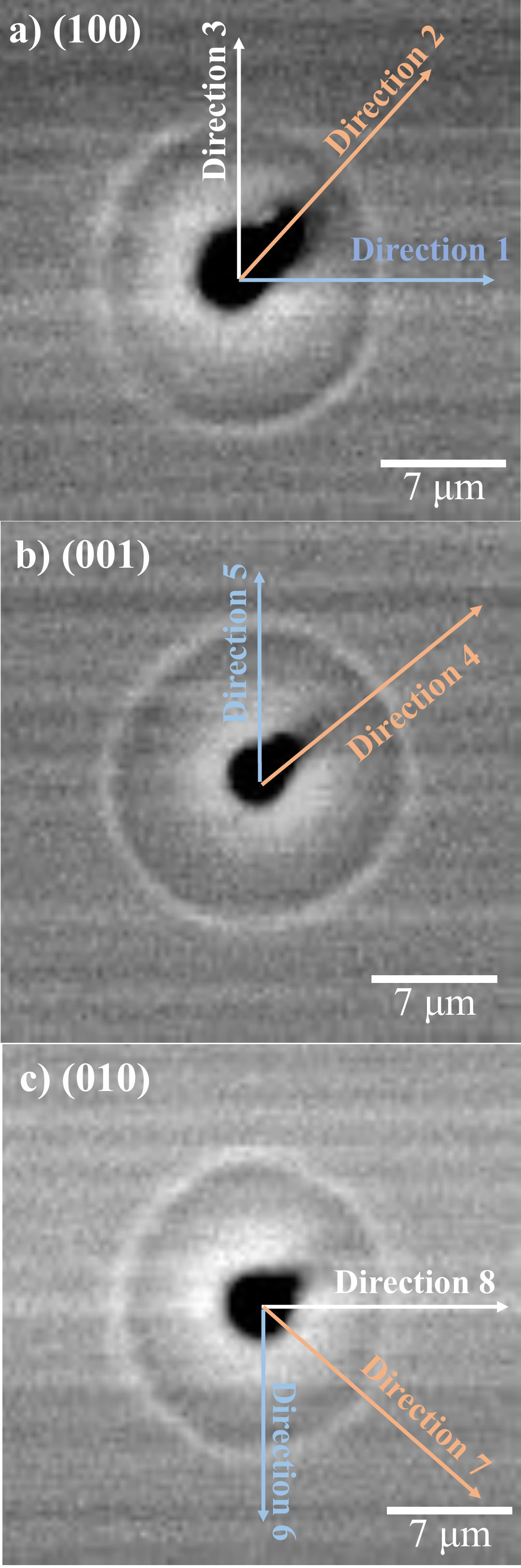}
		\caption{Scan of the TbMnO$_{3}$ (Pnma frame reference)  (a) (100), (b) (001) and (c) (010) surfaces at a fixed pump-probe delay of 3 ns showing the propagating acoustic waves (here shear ones only) emerging from the epicenter. The studied directions have been drawn.}
		\label{Fig1}
	\end{center}
\end{figure}

Figures~\ref{Fig1} (a, b, c) show the propagation of the acoustic waves on the (100), (001) and (010) \TB surface, respectively. The signal is recorded by scanning the surface of the sample (28$\times$28 $\mu$m$^2$) at fixed pump-probe delay. It corresponds to a snapshot taken after 3 ns from the coincidence (temporal overlap of the pump and probe). The black area in the center of the image is from  photothermal stationary component while the contrast in intensity corresponds to maxima and minima in the amplitude of the wavefront. For the present time- and length-scale, we see only in Fig.~\ref{Fig1} one wavefront with an elliptic shape for each planes identified later as a shear vertical wave. 

\begin{figure}[tb!]
	\begin{center}
		\includegraphics[width=5.5cm]{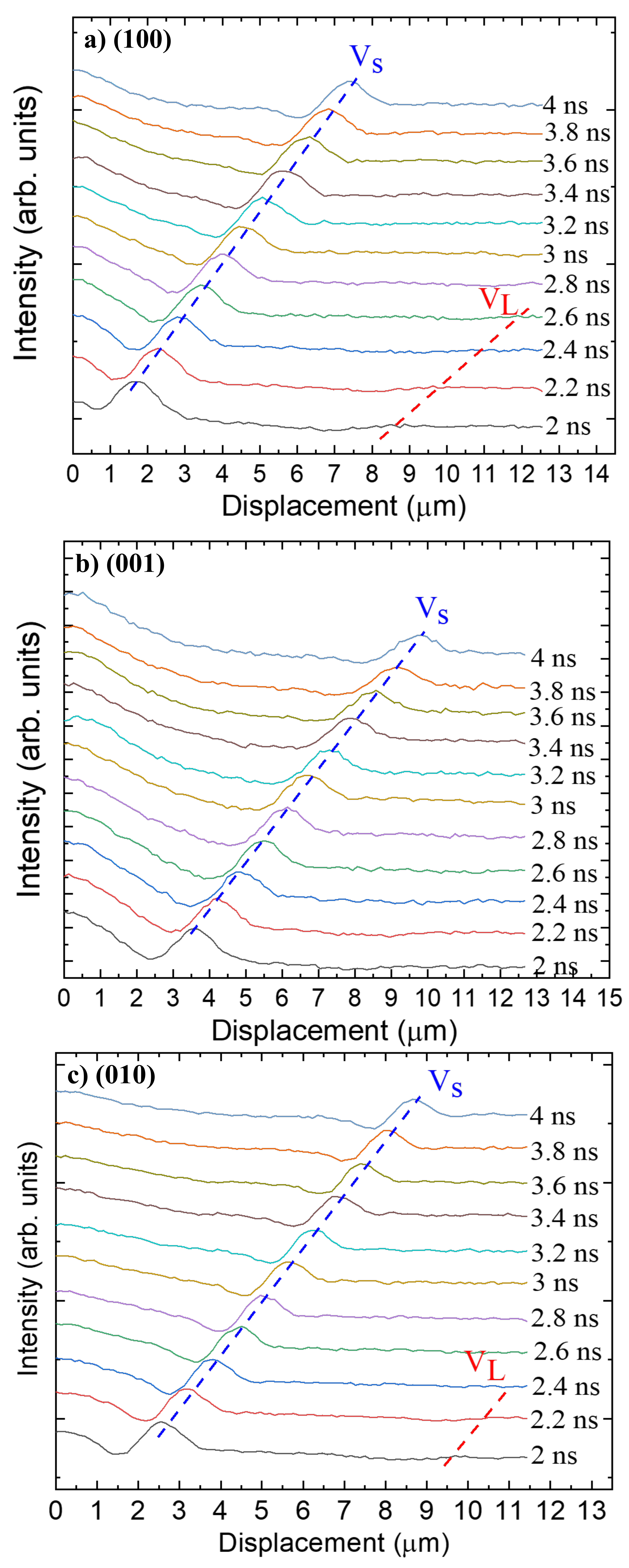}
		\caption{Relative phase change of the electromagnetic field of the probe beam as a function of the displacement and at different probe time delay, along analyzed directions:  (a) "2", (b) "5" and (c) "7".}
		\label{Fig2}
	\end{center}
\end{figure}

Hence, we have performed scan lines along at least two directions exhibiting highest and lowest propagating sound velocities. Profiles of the acoustic waves at different times of probe arrival after the coincidence, from 2 ns to 4 ns by step of 200 ps are shown in Figs. \ref{Fig2}(a, b, c). They present the relative change of phase of the probe reflectivity along the scanned "direction 2", "direction 5" and "direction 7" on (100), (001) and (010) plane, respectively, as a function of the probe position on the sample, along the given direction. One can observe in Figs. \ref{Fig2}, either, one or two peaks (see the  supplementary material, the second peak is associated to a second wavefront not observable in Fig. \ref{Fig1}) whose positions evolve as a function of the probe delay time, the first one being always much more intense. These peaks correspond to wave amplitude maxima and distinct propagation speeds, the first peak being the slowest one. For seek of simplicity, analysis will first consider bulk acoustic waves and later surface acoustic waves, all remaining close to each other. 
Our measurements are performed on (100), (001) and (010) planes, the three high symmetry planes of the orthorhombic crystal. In these planes, one longitudinal (L) and one shear (S1) acoustic waves are polarized and propagating parallel to the plane, and the second shear wave (S2) has polarization perpendicular to the plane (see description in the supplementary material). In total, eight directions were investigated, from which exact attribution and description of these peaks (Fig.\ref{Fig2}) could be performed thanks to analytical relations described in the supplementary material and are reported in Table \ref{Tab1}. In one plane, maximum and minimum sound velocity is necessarily along one high symmetry direction x$_i$ (i=1..3). The appearance of one common direction on each plane with similar sound velocity was considered, and helped us to identify highest symmetry directions x$_1$, x$_2$ and x$_3$. 

\begin{table} [h]    
	\begin{ruledtabular}
		\begin{tabular}{c|c|c|c}
			{Label} & {Velocity (L or S)} & {Direction(plane)} & {Elastic constants}   \\
			\hline
			1 & (L) 5150$\pm$300 & [010](100) & C$_{22}$ 		      \\ 
			1 & (S) 2733$\pm$22  & [010](100) & C$_{66}$ or C$_{44}$  \\ 
			2 & (L) 5359$\pm$82  & [011](100) & C$_{b+c, L}$		  \\ 
			2 & (S) 2820$\pm$16	 & [011](100) & $\frac{C_{66}+ C_{55}}{2}$ or C$_{b+c, S1}$ \\ 
			3 & (L) 4911$\pm$38  & [001](100) & C$_{33}$		      \\ 
			3 & (S) 2861$\pm$38  & [001](100) & C$_{55}$ or C$_{44}$  \\ 
			4 & (S) 2933$\pm$20  & [110](001) & $\frac{C_{55}+ C_{44}}{2}$ or C$_{a+b, S1}$	 \\
			5 & (S) 3070$\pm$15	 & [010](001) & C$_{44}$ or C$_{66}$  \\
			6 & (S) 3114$\pm$13  & [$\bar{1}$01](010) & $\frac{C_{66}+ C_{44}}{2}$ or C$_{a+c, S1}$	  \\ 
			7 & (L) 4841$\pm$150 & [001](010) & C$_{33}$              \\ 
			7 & (S) 3044$\pm$17  & [001](010) & C$_{44}$ or C$_{55}$  \\
			8 & (S) 2947$\pm$23  & [10$\bar{1}$](010) & $\frac{C_{66}+ C_{44}}{2}$ or	    C$_{a+c, S1}$     \\
		\end{tabular}
	\end{ruledtabular}
	\caption{Measured sound velocities (V) along the analyzed directions labelled (1-8), identified crystallographic directions and relations with the C$_{ij}$=$\rho$ V$^{2}$ elastic constants. V$_{L}$ and V$_{S}$ are the longitudinal and shear sound velocities given in m/s. C$_{ij}$ is related to the shear wave (S2) with a vertical (out-of-plane) polarization.\\ 
		C$_{b+c, L}$=$\frac{C_{22}+C_{33}+2C_{44}}{4}+\frac{1}{2}\sqrt{\frac{(C_{33}-C_{22})^{2}+4(C_{44}+C_{23})^{2}}{4}}$\\
		C$_{b+c, S1}$=$\frac{C_{22}+C_{33}+2C_{44}}{4}-\frac{1}{2}\sqrt{\frac{(C_{33}-C_{22})^{2}+4(C_{44}+C_{23})^{2}}{4}}$\\
		C$_{a+c, S1}$=$\frac{C_{11}+C_{33}+2C_{55}}{4}-\frac{1}{2}\sqrt{\frac{(C_{33}-C_{11})^{2}+4(C_{55}+C_{13})^{2}}{4}}$\\
		C$_{a+b, S1}$=$\frac{C_{11}+C_{22}+2C_{66}}{4}-\frac{1}{2}\sqrt{\frac{(C_{22}-C_{11})^{2}+4(C_{66}+C_{12})^{2}}{4}}$}
	\label{Tab1}
\end{table}

In order to attribute the measured sound velocities to the correct acoustic modes, we have implemented several calculation steps to optimize the elastic constants by minimizing the difference between the experimental and analytical theoretical velocities.
Knowing the C$_{ij}$ elastic constants and mass density (we used $\rho=$7603 kg/m$^{3}$) of the material, allows to calculate sound velocities by solving the Christoffel equation (section A in the supplemental material for any direction in (100), section B for the (010) plane and section C for the (001) plane). According to the \TB  space-group symmetry (Pnma), the elastic constant tensor C$_{ij}$ has nine independent constants, \cite{Born} namely, C$_{11}$, C$_{12}$, C$_{13}$, C$_{22}$, C$_{23}$, C$_{33}$, C$_{44}$, C$_{55}$ and C$_{66}$, sufficient to simulate any other elastic properties.
For ab initio calculations, we first followed a similar strategy than our previous work on BFO\cite{Pierre2021}. Electronic-structure calculations are carried out using a plane-wave pseudopotential approach to DFT as implemented within the VASP code \cite{Kresse1986,Kresse1986b} using the electron-ion interaction, described via the projector augmented wave method and potentials\cite{Bloch1994, Kresse1999}. We considered a collinear-spin magnetization type with generalized gradient approximation (GGA), to depict the exchange correlation functional. We used both the Perdew–Burke–Ernzerhof (PBE) and the one revised for solids (PBEsol) \cite{Perdew} to vary the equilibrium atomic volume at which the elastic properties are determine, with on-site Coulomb interactions (GGA+U). In this simplified approach of Dudarev {\it et al.} \cite{Dudarev1998}, we let Hubbard effective parameter U to vary between 0 and 7 eV to search for the band gap close to the experimental value of 0.5 eV, \cite{Cui2005} and spanning a large range of equilibrium volume and lattice parameters.  This is a spin-polarized magnetic calculation, with electronic iterations convergence of 0.0001 meV using the Normal (blocked Davidson) algorithm, plane wave cutoff energy of 520 eV and reciprocal space projection operators. The Brillouin-zone {\it k}-mesh is forced to be centered on the gamma point and corresponds to actual {\it k}-spacings of 6$\times$5$\times$6. The stress-strain method as implemented in VASP is employed (VASP-TAG IBRION=6), and elastic constants are determined on the relaxed structures by using the tetrahedron method incorporating Bl\"{o}chl corrections.\cite{Blochl1994} 

\begin{figure}[tb!]
	\begin{center}
		\includegraphics[width=8.5cm]{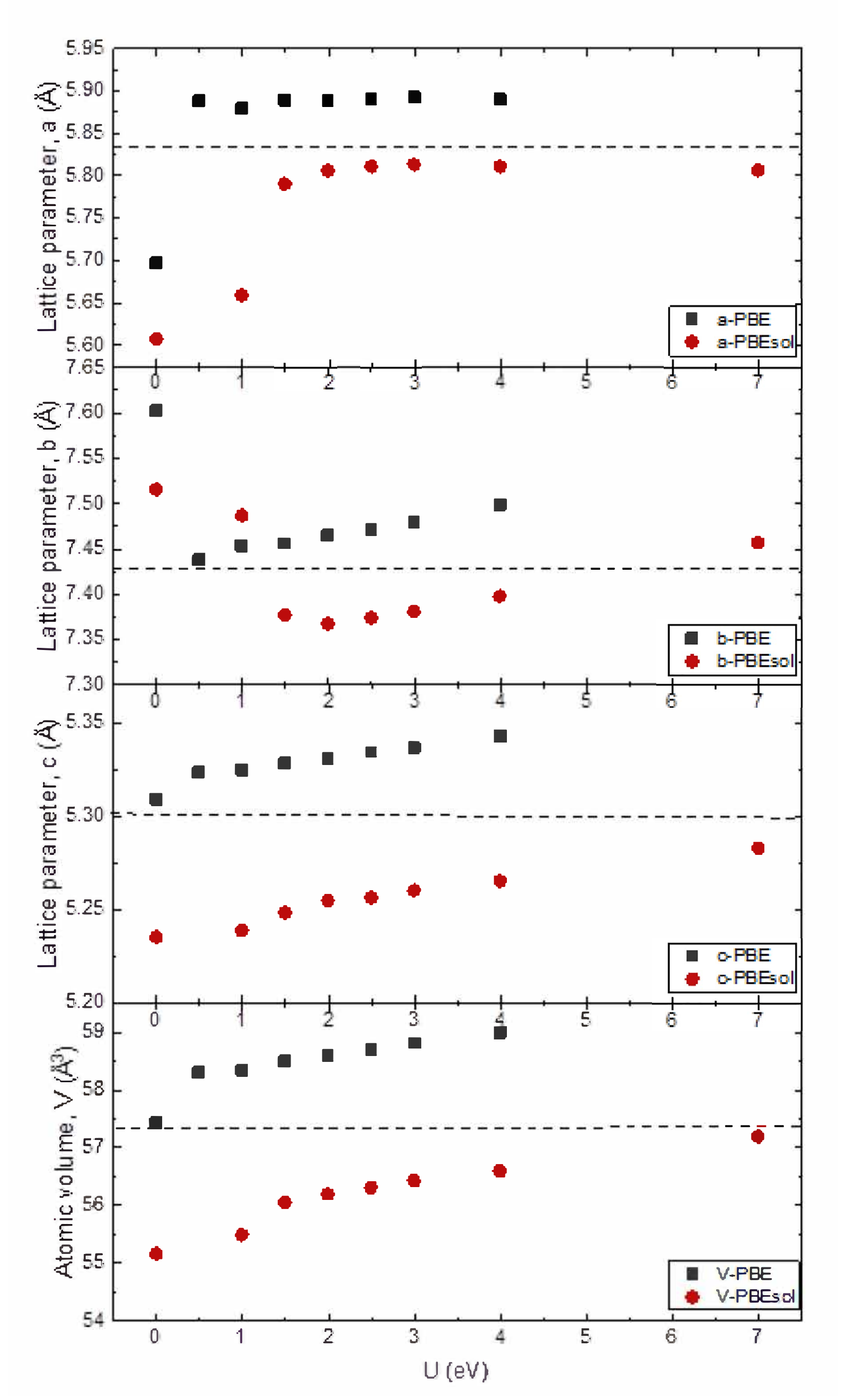}
		\caption{Lattice parameter and atomic volume calculated by GGA-PBE and GGA-PBEsol as a function of the effective Hubbard U term. Our experimental values are shown by a dashed-line.}
		\label{Fig3}
	\end{center}
\end{figure}

Due to approximations in the DFT+U method, we cannot directly compare energetics (total energy E$_t$) at different values of U. However we can compare energies that have been referenced to a common value, for example the ground state defined by the cell volume V$_0$(U) and the lowest total energy E$_0$(U), obtained for each U value. Rather than using this representation (E$_{real}$=E$_t$-E$_0$) versus the volume for the two flavor of exchange functional (GGA-PBE and GGA-PBESol), we kept the modification of the ground state atomic volume versus U, as shown in Fig.\ref{Fig3}. It has the advantage to clearly show that GGA-PBESol + U is approaching the experimental value V$_0$$\approx$57.19-57.43 $\AA^3$ for U$\approx$7 eV. The dynamical stability has been evaluated thanks to the phonon spectrum in the supplemental material.\\


All nine C$_{ij}$ values calculated by DFT are reported in Table \ref{Tab2}. 
While GGA-PBE is always overestimating, closer values of lattice parameters and cell volume are found for GGA-PBEsol with U=7 eV, while band gap is 0.529 eV. Then, first comparison with experimental sound velocities and elastic constants can be done with those obtained by this GGA-PBEsol+U=7 eV functional, refered later as "DFT" values. 

\begin{table} [h]    
	\begin{ruledtabular}
		\begin{tabular}{c|c|c|c|c|c}
			& {DFT$^a$} & {DFT$^b$} & {Theo.$^{\cite{Malashevich2008}}$} & {Exp.}$^{\cite{Alonso2000}}$ & {Exp. final set}$^c$\\
			\hline
			C$_{11}$ 						  & 185-209 & 174-249 & 170 & - & -   \\ 
			C$_{12}$						  & 131-147 & 141-173 &  & - & -   \\ 
			C$_{13}$						  & 93-117 & 109-118 &  & - & -   \\ 
			C$_{22}$						  & 255-277 & 268-301 & 150 & - & 202$\pm$20   \\ 
			C$_{23}$						  & 110-128 & 121-144 &  & - & 103$\pm$6   \\ 
			C$_{33}$           				  & 239-257  & 265-275 & 197  & - & 181$\pm$2   \\ 
			C$_{44}$						  & 87-91   &  43-76 & 48 & - &  71$\pm$5   \\
			C$_{55}$						  & 50-77   &  75-83 & 73  &  - &  60$\pm$2   \\
			C$_{66}$           				  & 62-75   &  100-101 & 77  &  - &  61$\pm$4   \\ 
			V$_{0}$           				  & 58.0-59.1 & 55.5-57.19 &  56.5&  57.19  & 57.43$\pm$0.19 \\ 
			a           					  & 5.7-5.9    & 5.6-5.81  &   5.817   &  5.8384  &  5.83$\pm$0.08
			\\ 
			b           					  &7.46-7.61   & 7.46-7.51  &    7.438  &  7.4025  &  7.43$\pm$0.05
			\\ 
			c          						  & 5.34      & 5.23-5.28  &  5.228    & 5.2931      &  5.30$\pm$0.05
			\\ 
			
		\end{tabular}
	\end{ruledtabular}
	\caption{C$_{ij}$ elastic constants determined in this work by DFT with GGA-PBE$^a$$+$U and GGA-PBEsol$^b$$+$U exchange-correlation functional and on-site Coulomb interaction (U = 0-7 eV) simplified approach of Dudarev {\it et al.} \cite{Dudarev1998} and our experiments.$^c$ Comparison to values calculated by interatomic shell model\cite{Choithrani2011} and x-ray diffraction for lattice parameters.\cite{Alonso2000} Elastic constants values are given in GPa, the equilibrium lattice parameters (a, b and c) in {\AA} and atomic volume V$_0$ in {\AA}$^3$/atom. Notice that DFT+U has an impact on elastic properties through the ground state volume modification when varying U.}
	\label{Tab2}
\end{table}

Here are examples to illustrate our approach. We measured very closed sound velocity of L-wave (4911 and 4841 m/s) along direction x$_2$ in both (100) and (001) planes, as well as, very closed sound velocity of S-wave (3070 and 3044 m/s) along direction x$_3$ (direction 5) and x$_2$ (direction 7), in both (001) and (010) planes. The sound velocity of L-wave is associated to the C$_{22}$ elastic constant.
Remember that our interferometric detection scheme favours the observation of shear wave (S2) having vertical out-of-plane component polarization. The C$_{44}$ elastic constant is thus associated to the sound velocity of S-wave.
The C$_{66}$, $C_{55}$ and C$_{33}$ were determined with a second direction in the (100) plane. 
C$_{44}$ agrees very well with DFT one, while other measured elastic constants are below by 15-40\%. 

The determination of the remaining diagonal elastic constant C$_{11}$ need a second independent measurement of C$_{22}$ and C$_{33}$. Both constants were  determined by a complementary time domain Brillouin scattering (TDBS) experiment (see the supplementary material)  which measured the product of refractive index by the longitudinal sound velocity ($nV_{L}$) traveling along the normal of sample plane. This experiment allows mainly a study of out-of-plane modes compared to the acoustical pump-probe method which allows to probe mainly the in-plane modes. 
We calculated the anisotropic refractive index $n$ ($n_a$=2.16 and $n_c$=2.24) at a wavelength of $\approx$800 nm (photon energy 1.5 eV) to serve as one input for analyzing the results ($nV_{L}$) of TDBS experiments\cite{Raetz2019,Kuriakose2017} conducted for each planes.
To do so, we performed additional DFT calculations with the hybrid functional (Heyd–Scuseria–Ernzerho,  HSE06)\cite{Krukau2006, Paier2006, Kim2009} as implemented in VASP. The frequency-dependent optical properties are calculated using the VASP-TAG (LOPTICS=.TRUE.) and the complex shift used to smoothen the real part of the dielectric function is set to 0.1. Here we use the only one available experimental value of refractive index ($n_c$=2.35) measured at 30K,\cite{Trepakov2016} nearly 10\% higher than our DFT ones. 

The measured longitudinal sound velocities measured by TDBS and the deduced C$_{11}$, C$_{22}$ and C$_{33}$ elastic constants ($C=\rho V^2$) are reported in Table \ref{Tab3}. They show a reasonable agreement within 10\%, 25\% and 3\% with DFT ones from Table \ref{Tab2}, respectively (for a discussion of the agreement see the supplementary material). Notice that the TDBS measurements do not show transverse waves in TbMnO$_3$ contrary to the multiferroic BiFeO$_3$  compound\cite{Ruello2012, Pierre2021} which remains a special case.  

\begin{table} [h]    
	\begin{ruledtabular}
		\begin{tabular}{c|c|c}
			{V{$_L$} (m/s)} & {Direction} & {Elastic constant (GPa)} \\
			\hline
			5468$\pm$110 &  [100] & C$_{11}$=227$\pm$10  \\ 
			6780$\pm$134 &  [010] & C$_{22}$=349$\pm$17  \\ 
			6002$\pm$120 &  [001] & C$_{33}$=274$\pm$10  \\ 
		\end{tabular}
	\end{ruledtabular}
	\caption{Longitudinal sound velocity (V$_L$) and elastic constants (C$_{ii}$) determined  in this work by TDBS along the normal of each plane samples by TDBS. A common refractive index n$_b$=2.35 (photon energy, 1.5 eV) measured at 30K by Trepakov et $\it al.$ \cite{Trepakov2016} is used.}
	\label{Tab3}
\end{table}

\begin{figure}[tb!]
	\begin{center}
		\includegraphics[width=8cm]{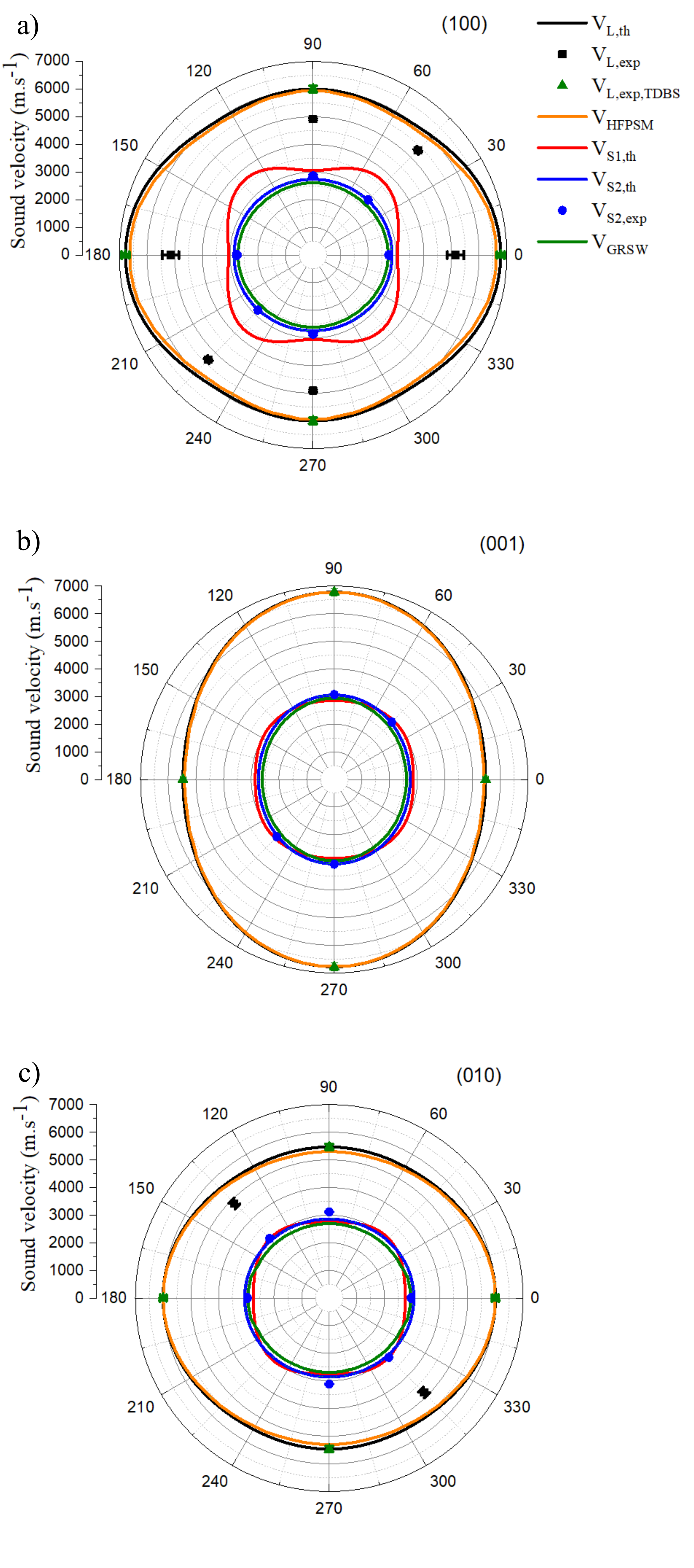}
		\caption{Comparison of the TbMnO$_{3}$ (Pnma frame reference)  (a) (100), (b) (010) and (c) (001) surfaces between experimental and simulated data.}
		\label{Fig4}
	\end{center}
\end{figure}

To push further the analysis, we performed the calculations of surface acoustic waves sound velocities which can propagate in each three high symmetry planes, considering the following set of elastic constants: C$_{11}$= 227 GPa, C$_{22}$= 349 GPa, C$_{33}$= 274 GPa, C$_{44}$= 71 GPa, C$_{55}$= 57 GPa, C$_{66}$= 62 GPa, C$_{12}$= 141 GPa,  C$_{13}$= 109 GPa, C$_{23}$= 103 GPa,\cite{Pierre2021} and compared them to the bulk ones in Fig. \ref{Fig4}.

Figure \ref{Fig4} shows the calculated and measured acoustic waves sound velocities on the (a) (100), (b) (001) and (c) (010) surfaces. We clearly see in this figure, first, the elliptical shape as in Fig.\ref{Fig1} and that the generalized Rayleigh surface acoustic wave (GRSW) with the lowest velocity remains very close to the bulk vertical shear wave ($S_2$) for all directions. The same observation is valid for the high frequency pseudo surface wave (HFPSW) and bulk longitudinal wave (L). This is a validation of our simplified approach with analytical expressions offered in case of bulk waves we used for the data treatment. However, for the in-plane measurements, we can't conclude on the nature of the wave, bulk or surface waves.

Notice that because it is not obvious to determine what kind of acoustic waves is propagating without any apriori knowledge of their sound velocity and polarization, we have performed preliminary DFT theoretical estimates to guide the identification. This helped us to distinguish between the faster ($\approx$5000 m/s) longitudinal wave (L) and the two transverse waves (S1 and S2) propagating with a lower velocity ($\approx$3000 m/s), by considering first the range of their respective sound velocity. In a second step, due to our detection schema, we privileged the observation of the only transverse wave having a vertical component (S2).

\begin{table} [h]   
	\begin{tabular}{|c|c|c|c|}
		\hline
		\multicolumn{4}{|c|}{Magnetostriction/Magnetoelastic coefficents}\\
		\hline	
		
		TbMnO$_3$ &    TbMnO$_3$      & BiFeO$_3$ & BiFeO$_3$                     \\
		&  x$10^{-6}$(MPa)  &           & x$10^{-6}$/no coeff.(MPa)     \\
		
		\hline          
		$\lambda_1$/b$_1$& -1.33/0.3    & $\lambda^{\alpha1,2}$/b$_{21}$ & -34.40/6.52\\ 
		$\lambda_2$/b$_2$& -9.97/0.4    & $\lambda^{\alpha2,2}$/b$_{22}$ & 61.94/-3.99\\ 
		$\lambda_3$/b$_3$& -1.87/0.4    & $\lambda^{\gamma,1}$/b$_{3}$   & 58.36/-4.72\\ 
		$\lambda_4$/b$_4$& 18.56/2.9    & $\lambda^{\gamma,2}$/b$_{4}$   & 62.26/3.91\\ 
		$\lambda_5$/b$_5$& 2.46/0.3     & $\lambda^{1,2}$/b$_{14}$       & 0.81/-5.86\\
		$\lambda_6$/b$_6$& -5.27/0.2    & $\lambda^{2,1}$/b$_{34}$       & -81.96/0.69\\
		$\lambda_7$/b$_7$& -3.03/1.8    & & \\
		$\lambda_8$/b$_8$& 0.25/0.008   & & \\
		$\lambda_9$/b$_9$& 2.62/0.2     & & \\
		\hline
	\end{tabular}
	\caption{The nine anisotropic magnetostriction/magnetoelastic coefficents of TbMnO$_3$ labelled $\lambda_i$/b$_i$ (i=  1..9, using the convention in Ref. \onlinecite{Mason1954}) calculated by DFT (GGA-PBEsol+U= 7eV) in the spin-polarized colinear case. We compare to ones of BiFeO$_3$ belonging to space group (R3c), we calculated in the same conditions (GGA-PBESol) as in our previous work [Ref. \onlinecite{Pierre2021}].}
	\label{Tab4}
\end{table}

As mentioned in the introduction, the determination of the magnetostriction and magnetoelastic coefficients are important to be able to quantify the interactions that give rise to rare phenomena and excitations observed in TbMnO$_3$.
Let's just remember that the degree of magnetostriction can be measured by the magnetostriction coefficient $\lambda$, which is the ratio of the fractional change in length (also known as strain or the change in length divided by the original length) to the magnetization of the material. 
Magnetostriction and magnetoelastic coefficients are directly related to each other by elastic constants.
We calculated the nine anisotropic magnetostriction and magnetoelastic coefficients labelled, respectively, $\lambda_i$ and b$_i$ (i=  1..9, respecting the convention in Ref. \onlinecite{Mason1954}) using the scripts from MAELAS code \cite{Nieves2020a} and reported in Table \ref{Tab4}. These coefficients have been calculated in the colinear spin phase of TbMnO$_3$. 

These calculated values of the magnetostriction parameters could not be compared due to the lack of experimental data measured under the conditions of our calculations. In Ref. \onlinecite{Meier}, magnetostrictive properties have been calculated at 0 K for a different magnetic state compared to the one considered here.
We have chosen to compare the magnetostrictive and magnetoelastic constants of TbMnO$_3$ to the ones of the BiFeO$_3$ (BFO) material. BFO is very well-known multiferroic compound with amazing properties (conductive domain walls, low bandgap unlike usual ferroelectrics,  spectacular THz electromagnetic wave generation). Even for this very studied compound the magnetostrictive and magnetoelastic constants are not well determined. Using the same scripts as the one used for TbMnO$_3$, the 6 anisotropic magnetostriction and magnetoelastic coefficients of BFO are reported in Table \ref{Tab4}.
	The magnetostriction values of TbMnO$_3$ obviously do not make this compound suitable for this type of applications. However, in motor shielding, electric transformers or magnetic recording, magnetic materials with extremely small magnetostrictive coefficients are useful. On the contrary, BFO presents large magnetostrictive coefficients needed for many applications in electromagnetic microdevices as actuators and sensors. Our calculated values will serve as a guide to the design or analysis of future experimental works.


In conclusion, acoustical pump-probe experiment at the nanosecond time-scale have been implemented to measure sound velocities of acoustic modes in single crystals TbMnO$_3$. Among the nine independent C$_{ij}$ elastic constants, seven  have been determined. In addition, anisotropic magnetostriction and magnetoleastic coefficients has been simulated by DFT.

\subsection*{SUPPLEMENTARY MATERIAL}
See supplementary material for the equations connecting the sound velocities of bulk waves propagating within the three high symmetry planes, to the C$_{ij}$ elastic constants. Complementary results obtained by time domain Brillouin scattering along each high symmetry directions $[100]$ are also provided.
\subsection*{ACKNOWLEDGMENTS}
The authors acknowledge J. Rastikian and S. Suffit for the evaporation of aluminium thin film on the \TB surface at the cleanroom of Université de Paris. 



\newpage
\newpage

	\section{{Details of the experimental setup}}
	
	\begin{figure}
		\begin{center}
			\includegraphics[width=7.5cm]{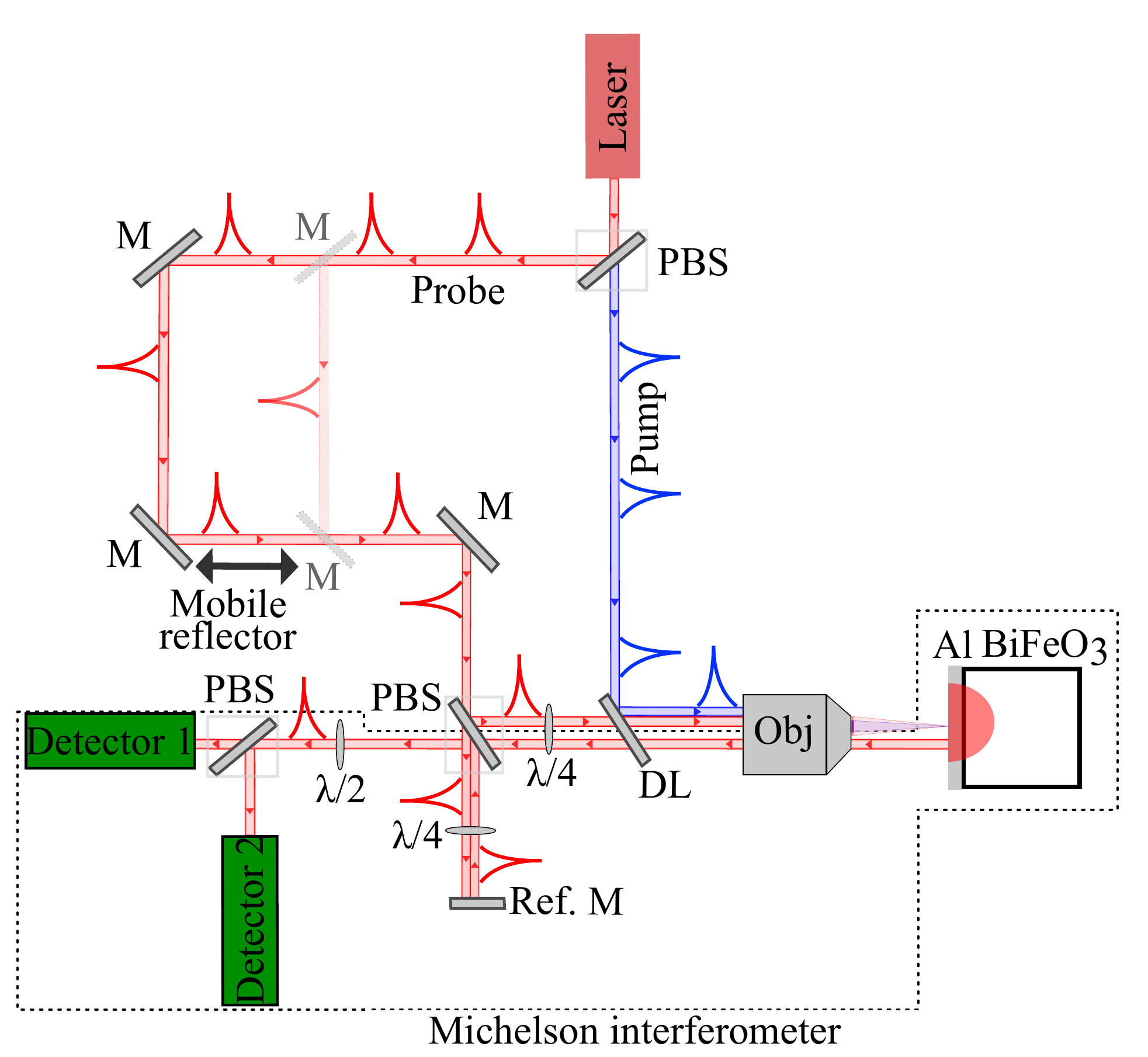}
			\caption{Sketch of the experimental setup with first the pump focused (obj is an objective) on TbMnO$_3$ capped with an Al thin film, second the probe path (PBS polarized beam splitters, M are mirrors) and the mobile reflector for the time delay, third the detection path and the Michelson interferometer to detect the perpendicular surface displacements (two detectors are used to servo-control the Michelson and also to measure the phase of the reflectivity change).}
			\label{Fig1}
		\end{center}
	\end{figure}
	
	For our pump‐probe experiment as shown in Fig.\ref{Fig1}, a mode-locked Ti:sapphire (MAI TAI Spectra) laser source operating at 800 nm was used to thermally excite and optically detect the propagating acoustic modes in TbMnO$_3$. The pulse duration is 200 fs and the pulse repetition rate is 80 MHz. 
	In order to get some above band gap excitation, the TbMnO$_3$ sample was excited by light obtained from second harmonic
	generation (SHG) with wavelength 400 nm performed by doubling the pump frequency with a nonlinear crystal (BBO). The power of the two beams is fixed at 300 $\mu$W. 
	Acoustic measurements were performed using a standard stabilized Michelson interferometer sensitive to the perpendicular surface displacement. A 12 ns maximum pump-probe time delay is achieved using a mobile reflector mounted on a translation stage. Both the pump and probe beams are focused with a microscope objective with a numerical aperture of 0.9, fixed on a piezoelectric stage. A tilting system of the probe allows the mapping of acoustic waves propagating in any direction, thus providing information on the elastic properties of TbMnO$_3$ single crystals. \
	
	\section{{Measurement of the longitudinal velocity V$_L$}}
	
\begin{figure}
		\begin{center}
			\includegraphics[width=5.5cm]{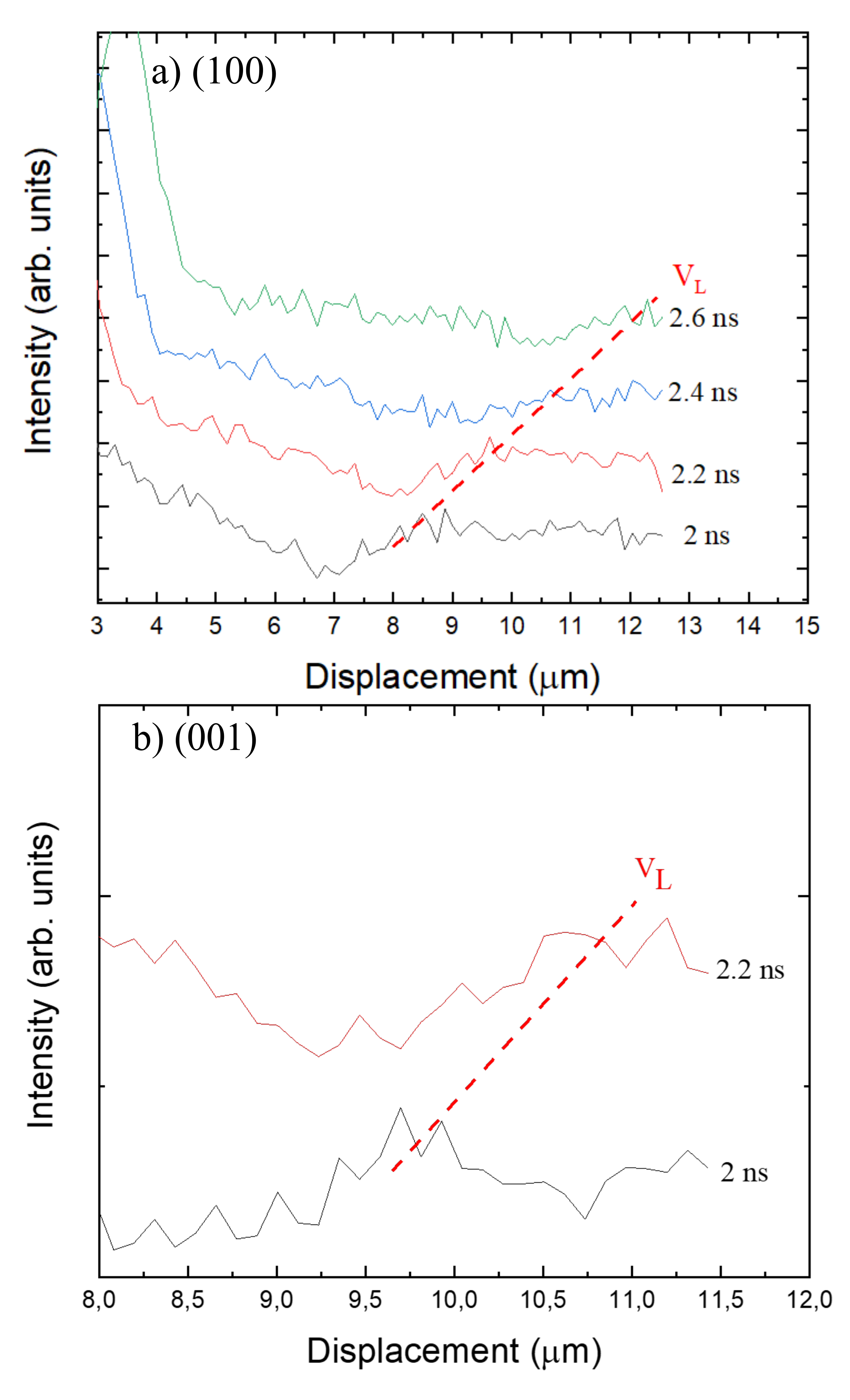}
			\caption{Relative phase change of the electromagnetic field of the probe beam as a function of the displacement and at different probe time delay, along analyzed directions:  (a) "2" and (b) "7".}
			\label{Fig2zoom}
		\end{center}
	\end{figure}

	Figure \ref{Fig2zoom} represents an enlargement of Fig. 2 in the article over the range of the longitudinal velocity V$_L$. 
	
	\section{{Bulk waves sound velocities by solving the Christoffel equations}}
	
	We provide first the equations connecting the sound velocities of bulk waves to the C$_{ij}$ elastic constants for each high symmetry planes [x$_2$,x$_3$], [x$_1$,x$_3$] and [x$_1$,x$_2$].
	
	In the Pnma representation ({\bf a}, {\bf b}, {\bf c}) of TbMnO$_3$ with angles $\alpha=\beta=\gamma=90\degree$, the symmetry of the elastic constant tensor is orthorhombic with 9 independent elastic constants:
	
	\begin{equation}
	C_{{\bf a}, {\bf b}, {\bf c}} = 
	\begin{pmatrix}
	C_{11} & C_{12} & C_{13} & 0 & 0 & 0 \\
	C_{12} & C_{22} & C_{23} & 0 & 0 & 0 \\
	C_{13} & C_{23} & C_{33} & 0 & 0 & 0 \\
	0 & 0 & 0 & C_{44} & 0 & 0 \\
	0 & 0 & 0 & 0 & C_{55} & 0 \\
	0 & 0 & 0 & 0 & 0 & C_{66} \\
	\end{pmatrix}
	\end{equation}
	
	The propagation of a bulk acoustic wave is described by Christoffel's equation :
	\newpage
	\begin{widetext}
	\begin{equation}\label{Cij}
	\rho\frac{\partial^2\vec{u}}{\partial t^2}=\vec{\nabla}\sigma
	\end{equation}
	
	with $\rho$ the volumic mass, $\vec{u}=\vec{u}_0e^{i(\vec{k}\vec{r}-\omega t)}$ the displacement vector where $\vec{k}$ is the wave vector, $\sigma_i$ is the stress tensor equal to $C_{ij}\epsilon_j$, $\epsilon$ being the deformation tensor. The phase speed is defined by : V=$\frac{\omega}{k}$.\\
	All following analytical results were checked by a fully numerical resolution using the program "Christoffel" provided in Refs. \onlinecite{JAEKEN2016a,Jaeken2016b} and agree with relations derived by Tsvankin.\cite{Tsvankin1997}
	
	Equation \ref{Cij} can be written in the following form:
	
	\begin{equation}\label{Ga}
	\begin{pmatrix}
	u_{1}  \\
	u_{2}  \\
	u_{3}  \\
	\end{pmatrix}
	= 
	\begin{pmatrix}
	C_{11}n^2_1 + C_{66}n^2_2 + C_{55}n^2_3 -\rho {V^2}& (C_{12} + C_{66})n_1n_2 & (C_{13} + C_{55})n_1n_3 \\
	(C_{12} + C_{66})n_1n_2 & C_{66}n^2_1 + C_{22}n^2_2 + C_{44}n^2_3 -\rho {V^2}& (C_{23} + C_{44})n_2n_3 \\
	(C_{13} + C_{55})n_1n_3 & (C_{23} + C_{44})n_2n_3 & C_{55}n^2_1 + C_{44}n^2_2 + C_{33}n^2_3 -\rho {V^2} \\
	\end{pmatrix}
	\end{equation}

	Here, $\vec{n}$ is the unit vector in the slowness direction.
	
	\subsection{Propagation in the [x$_2$,x$_3$] plane}
	
	If the wave vector lies in the [x$_2$,x$_3$] plane, we can define $\vec{n}(0,sin(\phi),cos(\phi))$ with respect to the {\bf(c)} direction ($\phi=0$). Then, equation \ref{Ga} now reads:

	\begin{equation}\label{G23}
	\begin{pmatrix}
	u_{1}  \\
	u_{2}  \\
	u_{3}  \\
	\end{pmatrix}
	= 
	\begin{pmatrix}
	C_{66}n^2_2 + C_{55}n^2_3 -\rho {V^2}& 0 & 0 \\
	0 &  C_{22}n^2_2 + C_{44}n^2_3 -\rho {V^2}& (C_{23} + C_{44})n_2n_3 \\
	0 & (C_{23} + C_{44})n_2n_3 &  C_{44}n^2_2 + C_{33}n^2_3 -\rho {V^2}\\
	\end{pmatrix}
	\end{equation}

	The solutions are for propagation in  the  [x$_2$,x$_3$] plane:
	
	1. a quasi-longitudinal mode polarized along (u$_2$,u$_3$)  with speed equal to V$_{L}$ = V$_+$,\\
	
	2. a quasi-shear mode polarized along (u$_2$,u$_3$) and perpendicular to the quasi-longitudinal one, with speed equal to V$_{S1}$ = V$_-$:

	\begin{equation}
	\rho {V^2_{+/-}}=\frac{b \pm \sqrt{b^2-4c}}{2} 
	\end{equation}\\
	with 
	\begin{equation}
	b=(C_{44}+C_{33})cos^2(\phi) +(C_{44}+C_{22})sin^2(\phi)
	\end{equation}

	\begin{equation}
	c=-[(C_{44}+C_{23})cos(\phi) sin(\phi)]^2+(C_{44}cos^2(\phi)+C_{22}sin^2(\phi))(C_{33} cos^2(\phi)+C_{44}sin^2(\phi))
	\end{equation}

	3. a pure shear mode vertically polarized along (u$_1$) with a velocity V$_{S2}$:
	\begin{equation}
	\rho {V^2_{S2}}={C_{66}sin^2(\phi)+C_{55}cos^2(\phi)}
	\end{equation}
	
		\subsection{Propagation in the [x$_1$,x$_3$] plane}
	If the wave vector lies in the [x$_1$,x$_3$] plane, we can define $\vec{n}(sin(\phi),0,cos(\phi))$ with respect to the {\bf(c)} direction ($\phi=0$). Then, equation \ref{Ga}  now reads:
	
	\begin{equation}\label{G13}
	\begin{pmatrix}
	u_{1}  \\
	u_{2}  \\
	u_{3}  \\
	\end{pmatrix}
	= 
	\begin{pmatrix}
	C_{11}n^2_1 + C_{55}n^2_3 -\rho {V^2}& 0 & (C_{13} + C_{55})n_1n_3\\
	0 &C_{66}n^2_1 + C_{44}n^2_3 -\rho {V^2}& 0 \\
	(C_{13} + C_{55})n_1n_3 & 0 &  C_{55}n^2_1 + C_{33}n^2_3 -\rho {V^2}\\
	\end{pmatrix}
	\end{equation}

	The solutions are for propagation in  the  [x$_1$,x$_3$] plane:\\
	
	1. a quasi-longitudinal mode polarized along (u$_1$,u$_3$)  with speed equal to V$_{L}$ = V$_+$,\\
	
	2. a quasi-shear mode polarized along (u$_1$,u$_3$) and perpendicular to the quasi-longitudinal one, with speed equal to V$_{S1}$ = V$_-$:
	\begin{equation}
	\rho {V^2_{+/-}}=\frac{b \pm \sqrt{b^2-4c}}{2} 
	\end{equation}
	\\
	with 
	\begin{equation}
	b=(C_{55}+C_{33})cos^2(\phi) +(C_{55}+C_{11})sin^2(\phi)
	\end{equation}
	\begin{equation}
	c=-[(C_{55}+C_{13})cos(\phi) sin(\phi)]^2+(C_{55}cos^2(\phi)+C_{11}sin^2(\phi))(C_{33} cos^2(\phi)+C_{55}sin^2(\phi))
	\end{equation}\\
	
	3. a pure shear mode vertically polarized along (u$_2$) with a velocity V$_{S2}$:
	\begin{equation}
	\rho {V^2_{S2}}={C_{66}sin^2(\phi)+C_{44}cos^2(\phi)}
	\end{equation}
	
	\subsection{Propagation in the [x$_1$,x$_2$] plane}
	If the wave vector lies in the [x$_1$,x$_2$] plane, we can define $\vec{n}(sin(\phi),cos(\phi),0)$ with respect to the {\bf(b)} direction ($\phi=0$). Then, equation \ref{Ga}  now reads:
	\begin{equation}\label{G12}
	\begin{pmatrix}
	u_{1}  \\
	u_{2}  \\
	u_{3}  \\
	\end{pmatrix}
	= 
	\begin{pmatrix}
	C_{11}n^2_1 + C_{66}n^2_2 -\rho {V^2}& (C_{12} + C_{66})n_1n_2 & 0\\
	(C_{12} + C_{66})n_1n_2 &C_{66}n^2_1 + C_{22}n^2_2 -\rho {V^2}& 0 \\
	0 & 0 &  C_{55}n^2_1 + C_{44}n^2_2 -\rho {V^2}\\
	\end{pmatrix}
	\end{equation}
	
	The solutions are for propagation in  the  [x$_1$,x$_2$] plane:\\
	
	1. a quasi-longitudinal mode polarized along (u$_1$,u$_2$)  with speed equal to V$_{L}$ = V$_+$,\\
	
	2. a quasi-transverse mode polarized along (u$_1$,u$_2$) and perpendicular to the quasi-longitudinal one, with speed equal to V$_{S1}$ = V$_-$:
	\begin{equation}
	\rho {V^2_{+/-}}=\frac{b \pm \sqrt{b^2-4c}}{2} 
	\end{equation}
	with 
	\begin{equation}
	b=(C_{66}+C_{22})cos^2(\phi) +(C_{66}+C_{11})sin^2(\phi)
	\end{equation}
	\begin{equation}
	c=-[(C_{66}+C_{12})cos(\phi) sin(\phi)]^2+(C_{66}cos^2(\phi)+C_{11}sin^2(\phi))(C_{22} cos^2(\phi)+C_{66}sin^2(\phi))
	\end{equation}
	
	3. a pure shear mode vertically polarized along (u$_3$) with a velocity V$_{S2}$:
	\begin{equation}
	\rho {V^2_{S2}}={C_{55}sin^2(\phi)+C_{44}cos^2(\phi)}
	\end{equation}
	
	\subsection{Propagation along the $<100>$ directions and time-domain Brillouin scattering experiments}
	
	For a wave vector and polarization along the [100], [010] and [001]  directions, the solutions are simple and enable selective determination of one elastic constant C$_{ii}$ (i=1..6):\\
	
	1. a shear mode with speed (V$_{S}$) such that $\rho$ V$^2_{S}$ = C$_{66}$ [100][010], C$_{44}$ [010][001], C$_{55}$ [001][100], first and second direction being the propagation and polarization direction, respectively;\\
	
	2. a longitudinal mode polarized according to (u$_i$) (i=1..3) with a velocity V$_L$ :
	\begin{equation}
	\rho {V^2_L}=C_{ii}
	\end{equation}\\
	
	\end{widetext}
	
	We have performed time-domain Brillouin scattering on TbMnO$_3$ single crystal along these three high symmetry directions. 
	
	\begin{widetext}	
	\begin{figure}[h]
		\begin{center}
			\includegraphics[width=18cm]{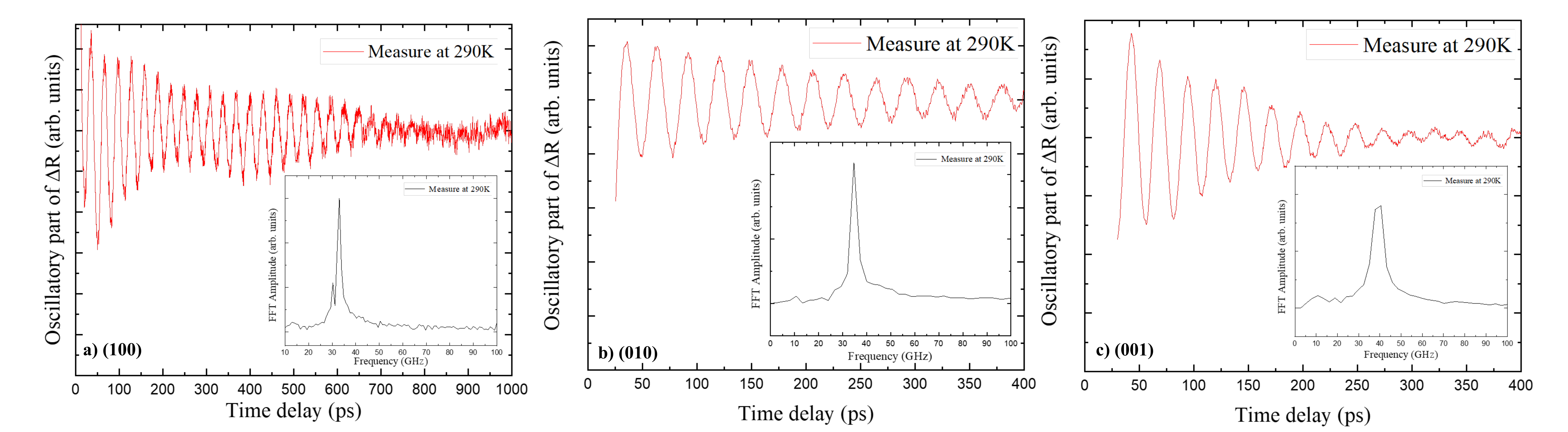}
			\caption{Fast Fourier transform of Brillouin oscillations detected along the three high symmetry directions.}
			\label{OS}
		\end{center}
	\end{figure}
	\end{widetext}
	
	Figure \ref{OS} shows the Fast Fourier transform of the oscillations observed in the time resolved optical reflectivity of (100), (010)  and (001) oriented single crystal (Pnma reference frame). For (100) investigated plane, one doublet is revealed at 30.25 GHz and at 32.97 GHz, measuring nV$_L$ for two distinct orthogonal polarization of the electric field having two different values of refractive index n, in the range [n$_b$,n$_c$]. They correspond to the  longitudinal (L) acoustic mode. The doublets arise from the birefringence in this (100) anisotropic plane, that we estimated to be n$_c$-n$_b$ $\approx$ 0.08) from our DFT calculations using hybrid HSEO6 functional. 
	
	In the back-scattering configuration used in our experiment, the momentum conservation is obtained for $\vec{q}_{ac}$=2$\vec{k}_{op}$ with $\vec{q}_{ac}$ the momentum of acoustic phonon and $\vec{k}_{op}$ the momentum of photon inside TbMnO$_3$. $k_{op}$=2$\pi n/\lambda$ where $\lambda$ is the wavelength of the laser probe and $n$ is the averaged refractive index of TbMnO$_3$ at the wavelength $\lambda$. The phonon dispersion law is assumed to be written as $2\pi f_{ac}$=V$q_{ac}$ where V is the sound velocity in TbMnO$_3$ associated to the Brillouin frequency $f_{ac}$. As an example, we can thus deduce the velocity of the longitudinal wave (V$_L$) along [100] direction from Fig. \ref{OS} a using $\lambda$=813 nm and our DFT calculated $n_b$=2.16 and $n_c$=2.26 and then C$_{11}$ elastic constant: 
	\begin{equation}
	V_L = 5693-5930\: m.s^{-1}   \hspace{10pt} C_{11}=257\pm10\:GPa
	\end{equation}
	
	One should notice that our refractive index values (Pnma representation) are lower than the experimental low temperature ($<$50K) refractive index $n_b$=2.35\cite{Trepakov2016} (photon energy 2.5 eV, $\lambda$=496 nm) and the theoretical one, $n_c$=2.7 derived by Lu {\it et al.}\cite{Lu2010} from DFT GGA+U(3eV).
	Contrary to our approach, the time-domain Brillouin scattering allows only to measure few directions around the normal direction and necessitates additional crystal with different orientations for further scrutinizing elastic properties.\\
	
	In Table III of the article, the measured longitudinal sound velocities measured by TDBS and the deduced C$_{11}$, C$_{22}$ and C$_{33}$ elastic constants show a reasonable agreement within 10\%, 25\% and 3\% with DFT ones from Table II, respectively.
	
	Nevertheless, it cannot be excluded that, if one consider the optical anisotropic character evidenced in ellipsometry measurements of Trepakov et $\it al.$,\cite{Trepakov2016} C$_{22}$ elastic constant reported in Table III is probably ~20 \% overestimated because of used ~10 \% undervalue of low temperature (30K) n$_a$ or n$_c$ (hence, sound velocity) in the TDBS experiment analysis. Indeed, n(T) dependence is expected to have the most general case of positive thermo-optic effect $(dn/dT > 0$) with higher refractive index values at room temperature (300K). Then, an overall agreement is plausible between out-of-plane and in-plane sound velocities, and elastic constants assessments. In the present work, these two approaches are found to be complementary as in-plane experiments did not evidenced propagating longitudinal wave within all high symmetry planes otherwise only a weak signal.
	\\Among the off-diagonal elastic constants, only C$_{23}$ could be measured thanks to longitudinal sound velocity along the bisecting direction [x$_2$, x$_3$] (direction 2) and agrees within 13\% with DFT one. Unfortunately, the longitudinal or the shear horizontal (S1)  waves were not observed for those bisecting directions [x$_1$, x$_3$] and [x$_1$, x$_2$] which are necessary to assess selectively the two last elastic constants C$_{13}$ and C$_{12}$.
	
	\section{{dynamical stability}}	
	
	\begin{widetext}
	In this part, we have studied the dynamical stability versus U parameter performing phonons spectrum calculations using GGA-PBESol exchange-correlation functional and on-site Coulomb interaction for four U values (0, 3.5, 7 and 9 eV). As shown in Fig.\ref{FigS4}, dynamical stability is expected above U=7 eV which exhibits two negative frequencies near the Brillouin-zone center, while phonons frequencies are all positive in case of U=9 eV. This result is in good agreement with the fact, that we reach experimental band gap and atomic volume for U$\approx$7eV and GGA-PBESol exchange-correlation functional.
	
		\begin{figure}[tb!]
		\begin{center}
			\includegraphics[width=10cm]{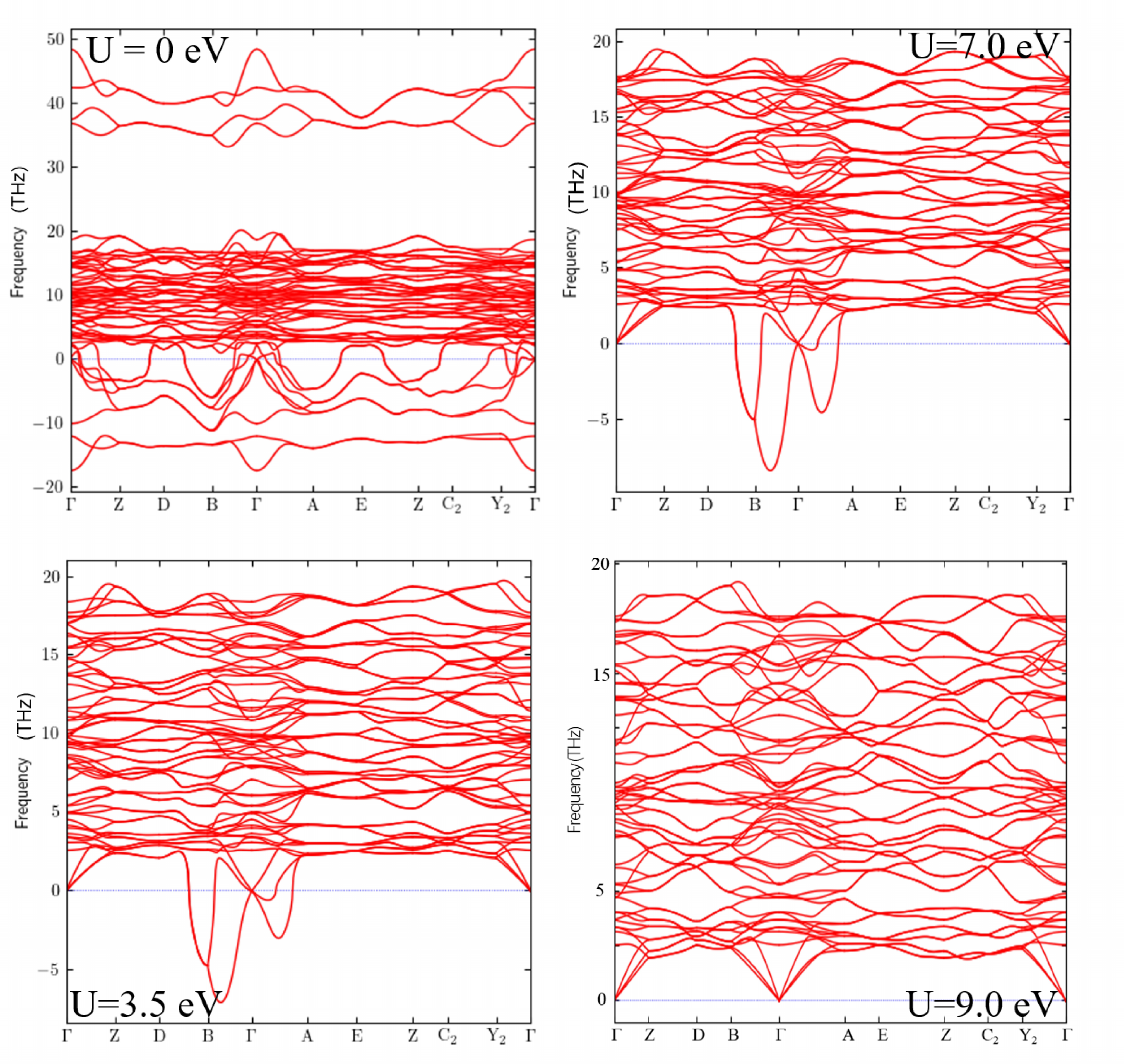}
			\caption{Phonons spectra of TbMnO$_3$ calculated for four on-site Coulomb interaction U (0, 3.5, 7 and 9 eV) using GGA-PBESol exchange-correlation functional.}
			\label{FigS4}
		\end{center}
	\end{figure}
	\end{widetext}


\end{document}